\documentclass{article}
\usepackage{graphicx}
\usepackage[english]{babel}

\title{
\includegraphics[width=0.35\textwidth]{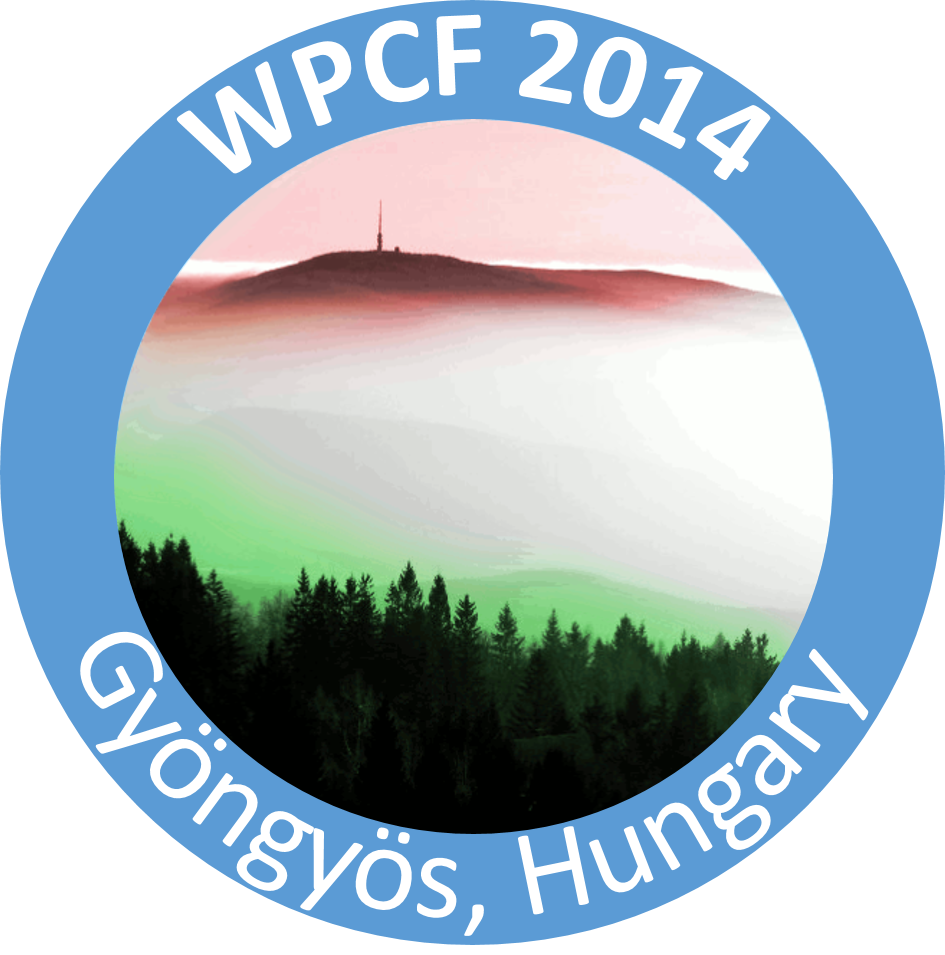}\\[1cm]
Two particle correlation effects and Differential HBT for rotation in heavy ion collisions. }
\author{L.P. Csernai, S. Velle and D.J. Wang\\[1ex]
$^1$Department of Physics and Technology, University of Bergen,\\ Allegaten 55, 5007 Bergen, Norway\\
}

\begin{document}

\fontfamily{lmss}\selectfont
\maketitle

\begin{abstract}
Peripheral heavy ion reactions at ultra relativistic energies 
have large angular momentum that can be studied via two particle 
correlations using the Differential Hanbury Brown and Twiss method. 
We analyze the possibilities and sensitivity of the method in a rotating system.  
We also study an expanding solution of the fluid dynamical
model of heavy ion reactions.
\end{abstract}

\section{Introduction}
Collective flow is one of the most dominant observable features
in heavy ion reactions up to the highest available energies, and
its global symmetries as well as its fluctuations are extensively
studied. Especially at the highest energies for peripheral reaction
the angular momentum of the initial state is substantial, which leads
to observable rotation according to fluid dynamical estimates
\cite{hydro1}.
Furthermore the low viscosity quark-gluon fluid may lead to
to initial turbulent instabilities, like the 
Kelvin Helmholtz Instability (KHI), according to numerical
fluid dynamical estimates
\cite{hydro2}, 
which is also confirmed in a simplified analytic model
\cite{KHI-Wang}. 
These turbulent phenomena further increase
the rotation of the system, which also leads to a large 
vorticity and circulation of the participant zone one order of 
magnitude larger than from random fluctuations 
in the transverse plane \cite{CMW12, FW11-1,FW11-2}.

The Differential Hanbury Brown and Twiss (DHBT) method 
has been introduced in \cite{DCF}. 
The method has been applied to a high resolution 
Particle in Cell Relativistic (PICR) fluid dynamical model \cite{CVW2014}. 

\section{The two particle correlation} \label{s2}
\label{PCF}

The pion correlation function is defined as the inclusive two-particle 
distribution divided by the product of the inclusive one-particle 
distributions, such that \cite{WF10}:
\begin{equation}
	C( p_1 , p_2) = 
	\frac{P_2( p_1, p_2)}{P_1( p_1)P_1( p_2)},
\end{equation}
where $ p_1$ and $ p_2$ are the 4-momenta of the pions 
and {k} and {q} are the average and relative momentum respectively. 

We use a method for
moving sources presented in Ref.\cite{Sinyukov-1}. 
In the formulae the $\hbar = 1$ convention is used and $k$ and $q$ are
considered as the wavenumber vectors.
The correlation function is:
\begin{equation}
	C(k,q) =
	1 + \frac{R(k,q)}{\left| \int d^4 x\,  S(x, k) \right|^2}\ ,
	\label{C-def}
\end{equation}
where
\begin{equation}
	R(k,q) = \int d^4 x_1\, d^4 x_2\, \cos[q(x_1-x_2)]  
	S(x_1,  {k}+{q}/2)\, S(x_2,{k}-{q}/2)\ .
	\label{R1}
\end{equation}
Here $R(k,q)$ can be calculated \cite{Sinyukov-1} via the function
and we obtain the $R(k,q)$ function as
\begin{equation}
	R(k,q) = {\rm Re}\, [ J(k,q)\, J(k,-q) ]
	\label{R-def}
\end{equation}
The corresponding $J(k,q)$ function will become
\begin{equation}
	J(k,q) =  \int d^4x\ S(x,k)\, 
	\exp\left[ - \frac{q \cdot u(x)}{2T(x)} \right]\, \exp(iqx)\ .
	\label{J2} 
\end{equation}

For the phase space distribution we frequently use the 
J\"uttner (relativistic Boltzmann) distribution, in terms of the local invariant
scalar particle density the J\"uttner distribution is \cite{Juttner}
\begin{equation}
	f^J(x,p) = \frac{n(x)}{C_n} 
	\exp\left(-\frac{p^\mu u_\mu (x)}{T(x)}\right)\ ,
	\label{Jut-2}
\end{equation}
where $C_n = 4 \pi m^2 T K_2(m/T)$. We assume a spatial distribution:

\begin{equation}
	G(x) = \gamma n(x) = 
	\gamma n_s  \exp\left( - \frac{x^2 + y^2 + z^2}{2 R^2} \right) .
\end{equation}
Here $n_s$ is the average density of the Gaussian source, $s$, (or fluid cell)
of mean radius $R$.

\textbf{Asymmetric Sources:} we have seen in few source model examples \cite{DCF}  that
a highly symmetric source may result in 
correlation functions that are sensitive to rotation, however, 
these results were not sensitive 
to the direction of the rotation, which seems to be unrealistic.
We saw that this result is a consequence of the assumption that
both of the members of a symmetric pair contribute equally to the
correlation function even if one is at the side of the system
facing the detector and the other is on the opposite side. 
The expansion velocities are also opposite at the opposite sides. The
dense and hot nuclear matter or the Quark-gluon Plasma are
strongly interacting, and for the most of the observed particle types
the detection of a particle from the side of the system, -- which
is not facing the detector but points to the opposite direction, --
is significantly less probable. The reason is partly in the diverging 
velocities during the expansion and partly to the lower emission
probability from earlier (deeper) layers of the source from the
external edge of the timelike (or spacelike) FO layer.

For the study of realistic systems where the emission is dominated
by the side of the system, which is facing the detector, 
we cannot use the assumption of the symmetry among pairs or 
groups of the sources from opposite sides of the system. Even if the
FO layer has a time-like normal direction, $\hat\sigma^\mu$ the 
$(k^\mu \hat\sigma^\mu)$ factor yields a substantial emission
difference between the opposite sides of the system. 

The correlation function, $C(k,q)$ is always measured
in a given direction of the detector, $\vec k$. Obviously
only those particles can reach the detector, which 
satisfy   $ \ k^\mu \hat\sigma_\mu \ > \ 0 $. Thus 
in the calculation of  $C(k,q)$ (see Fig. \ref{Fig-1}) for a given $\hat{\vec k}$-
direction we can exclude the parts of the freeze out
layer where $ \ k^\mu \hat\sigma_\mu \ < \ 0 $ (see Eq. (10)
of Ref.\cite{Sin89} or Ref.\cite{Bugaev}. For time-like 
FO a simplest approximation for the emission possibility is
$
P_{esc}(x) \ \propto \  k^\mu u_\mu(x) 
$
\cite{Cso-5}. 

\section{The DHBT method and fluid dynamical results}\label{EPFS}

Based on the few source model results the Differential HBT method \cite{DCF}
was introduced by evaluating the difference of two correlation
functions measured at two symmetric angles, forward and backward
shifted in the reaction plane in the participant c.m. frame
by the same angle, i.e. at $\eta = \pm $const., so that

\begin{equation}
	\Delta C(k,q) \equiv C(k_+,q_{out}) - C(k_-,q_{out}).
	\label{DCFdef}
\end{equation}

For the exactly $\pm x$ -symmetric spatial configurations
(i.e. $k_{+x} = k_{-x}$ and  $k_{+z} = - k_{-z}$), e.g. central collisions
or spherical expansion, $\Delta C(k,q)$ would vanish!
It would become finite if the rotation introduces an asymmetry.

The sensitivity of the standard correlation function on the 
fluid cell velocities decreases with decreasing distances
among the cells. So, with a large number of densely placed fluid cells
where all fluid cells contribute equally to the correlation function,
the sensitivity on the flow velocity becomes negligibly weak.

\begin{figure}[ht] 
	\vskip -0.3cm
	\begin{center}
		\includegraphics[width=7.6cm]{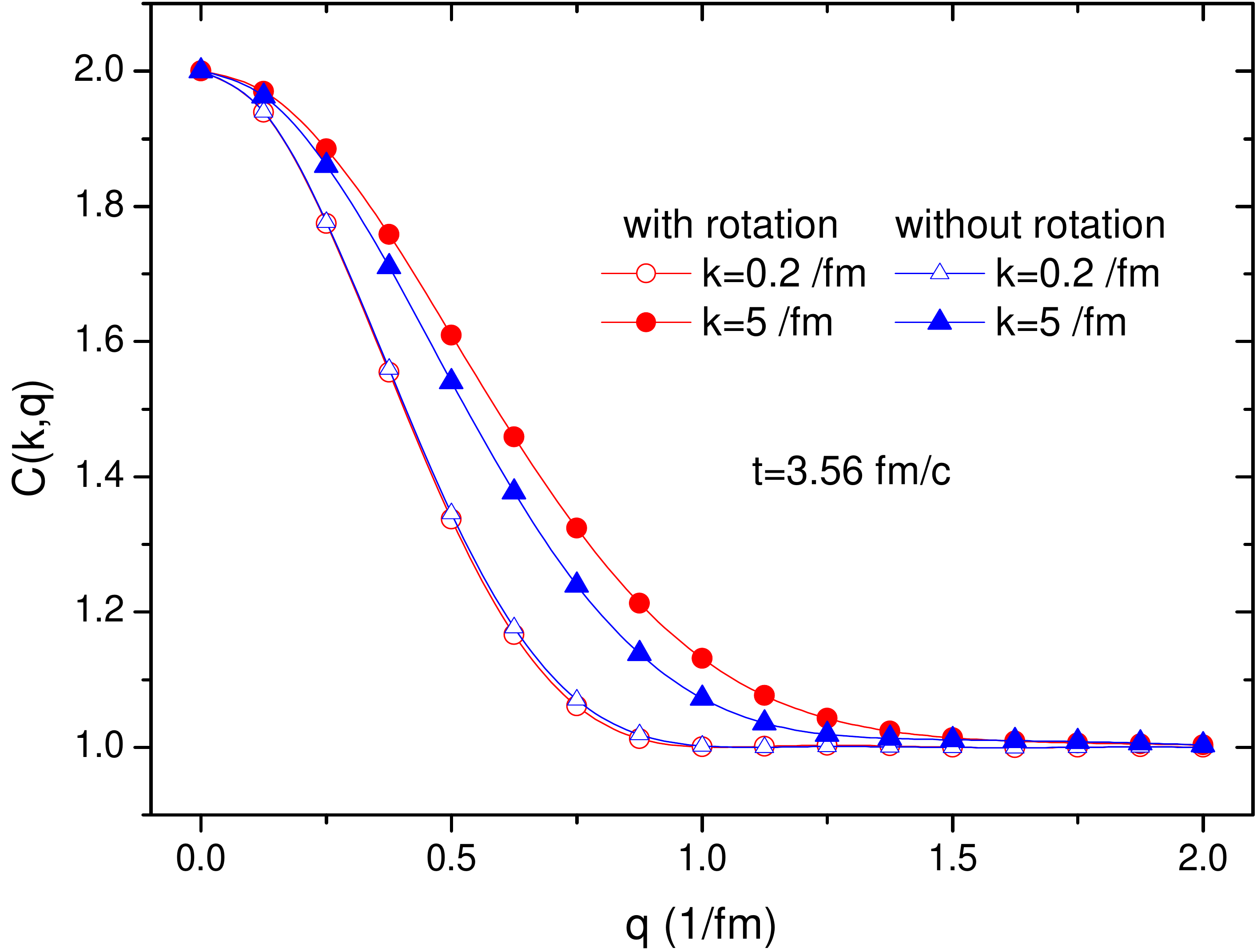}
	\end{center}
	\caption{(color online)
		The dependence of the standard correlation function in the $\vec k_+$
		direction from the collective flow, at the final time.
		From ref. \cite{CVW2014}.
	}
	\label{Fig-1}
\end{figure}

Thus, the emission probability from different ST regions of the
system is essential in the evaluation. This emission asymmetry
due to the local flow velocity
occurs also when the FO surface or layer is isochronous
or if it happens at constant proper time.

\begin{figure}[ht] 
	\vskip -0.3cm
	\begin{center}
		\includegraphics[width=7.6cm]{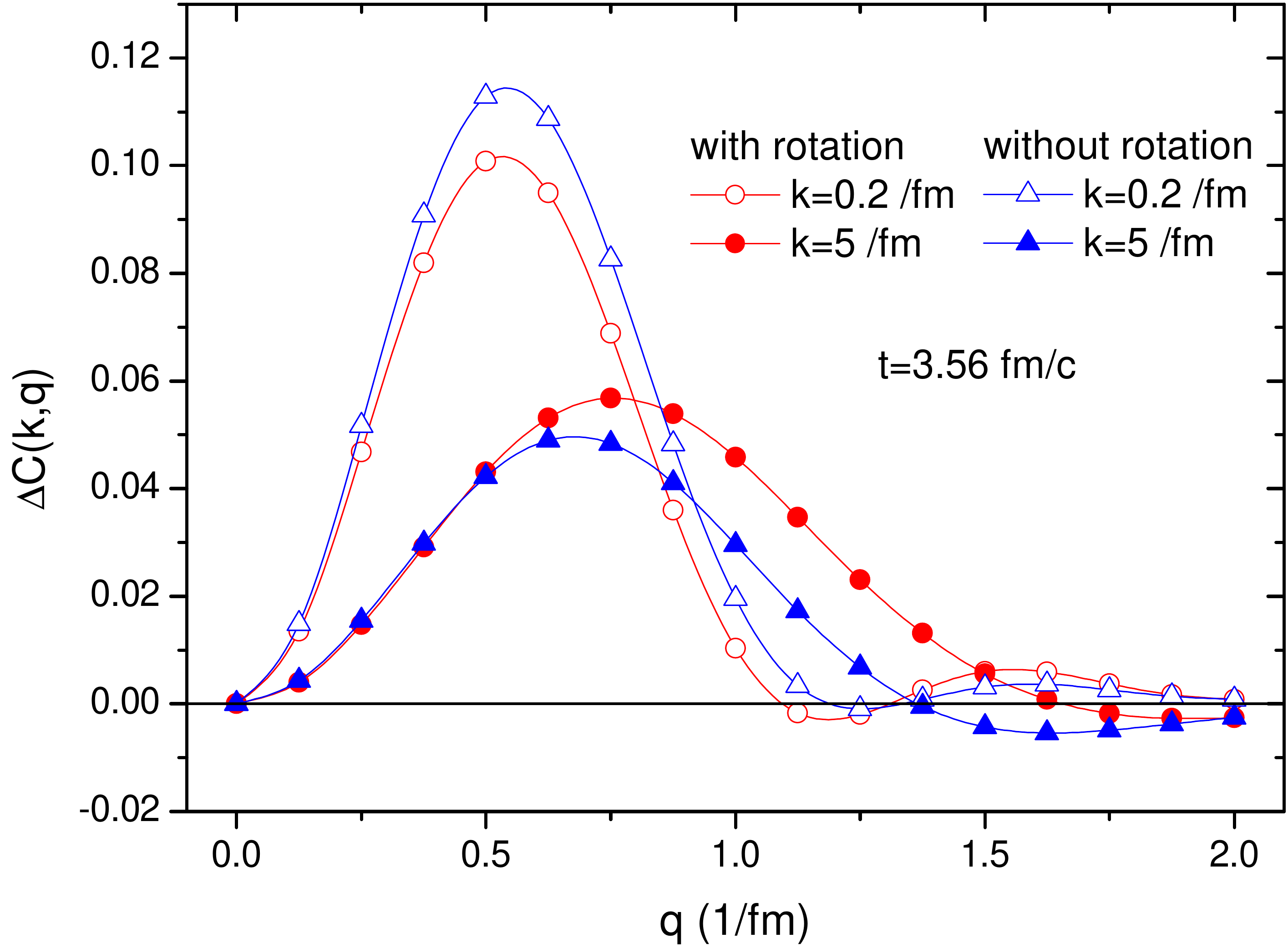}
	\end{center}
	\vskip -0.3cm
	\caption{(color online)
		The differential correlation function $\Delta C(k,q)$
		at the final time with and without rotation. 
		From ref. \cite{CVW2014}.
	}
	\label{Fig-2}
\end{figure}

We studied the fluid dynamical patterns of the calculations
published in Ref. \cite{hydro2},
where the appearance of the KHI is discussed under different conditions.
We chose the configuration, where both the rotation \cite{hydro1},
and {\bf the KHI occurred}, at
$b=0.7 b_{max}$ with high cell resolution and low numerical
viscosity at LHC energies, where the angular momentum is large,
$L \approx 10^6 \hbar$ \cite{VAC13}. 
Fig. \ref{Fig-2} shows the DHBT for the FD model.

The standard correlation function is both influenced by the
ST shape of the emitting source as well as its velocity
distribution. The correlation function becomes narrower in $q$
with increasing time primarily due to the rapid expansion of the
system. At the initial configuration the increase of $|\vec k|$
leads to a small increase of the width of the correlation function.

Nevertheless, in theoretical models we can switch off the rotation component
of the flow,
and analyse how the rotation influences the correlation function
and especially the DCF, $\Delta C(k,q)$.
\begin{figure}[ht] 
	\vskip -0.3cm
	\begin{center}
		\includegraphics[width=7.6cm]{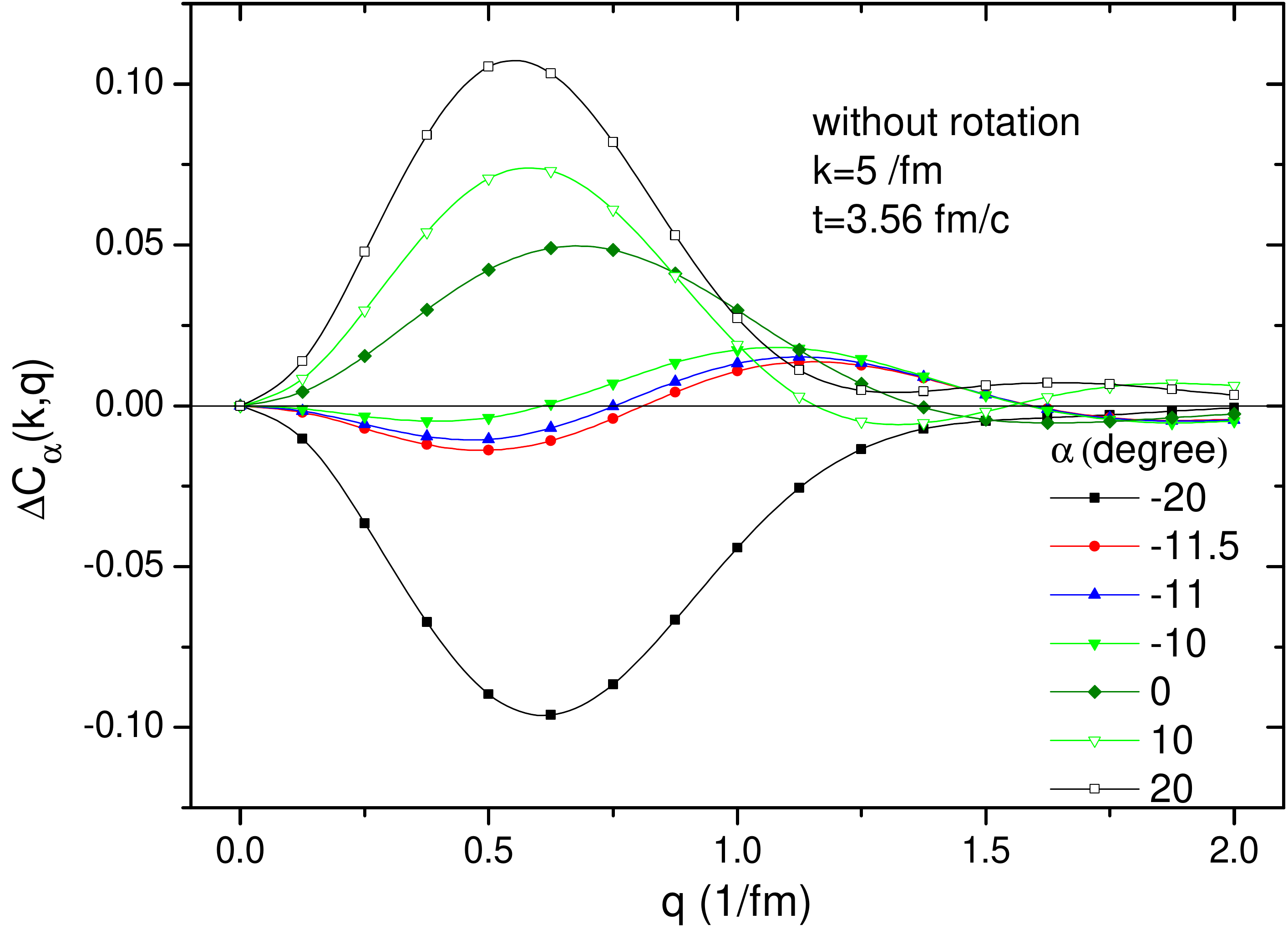}
	\end{center}
	\vskip -0.3cm
	\caption{(color online)
 		The Differential Correlation Function (DCF) at average pion wavenumber,
 		$k = 5/$fm and fluid dynamical evolution time, $t=3.56$fm/c, as a function
 		of the functions of momentum difference in the "out" direction $q$
 		(in units of 1/fm). The DCF is evaluated in a frame rotated in the
 		reaction plane, in the c.m. system by angle $\alpha$. From ref. \cite{CVW2014}.
	}
	\label{Fig-3}
\end{figure}

Fig. \ref{Fig-1} compares the standard correlation functions
with and without the rotation component of the  flow at the final time moment.
Here we see that the rotation leads to a small increase of the
width in $q$ for the distribution at high  values of $|\vec k|$, while at 
low momentum there is no visible difference.

In Fig.\,\ref{Fig-2}
$\Delta C(k,q) $ is shown for the configuration with and without
rotation. For $k=5/$fm the rotation increases both the 
amplitude  and the width of $\Delta C$. 
The dependence on $|\vec k|$ is especially large at the final time.

Fig. \ref{Fig-3} shows the result where the rotation component of the velocity
field is removed. The DCF shows a minimum in its integrated value
over $q$, for $\alpha = -11$ degrees.   The shape of the DCF changes 
characteristically with the angle $\alpha$. 
Unfortunately this is not possible experimentally, so the direction
of the symmetry axes should be found with other methods, like global flow 
analysis and/or azimuthal HBT analysis.
\begin{figure}[ht] 
	\vskip -0.3cm
	\begin{center}
		\includegraphics[width=7.6cm]{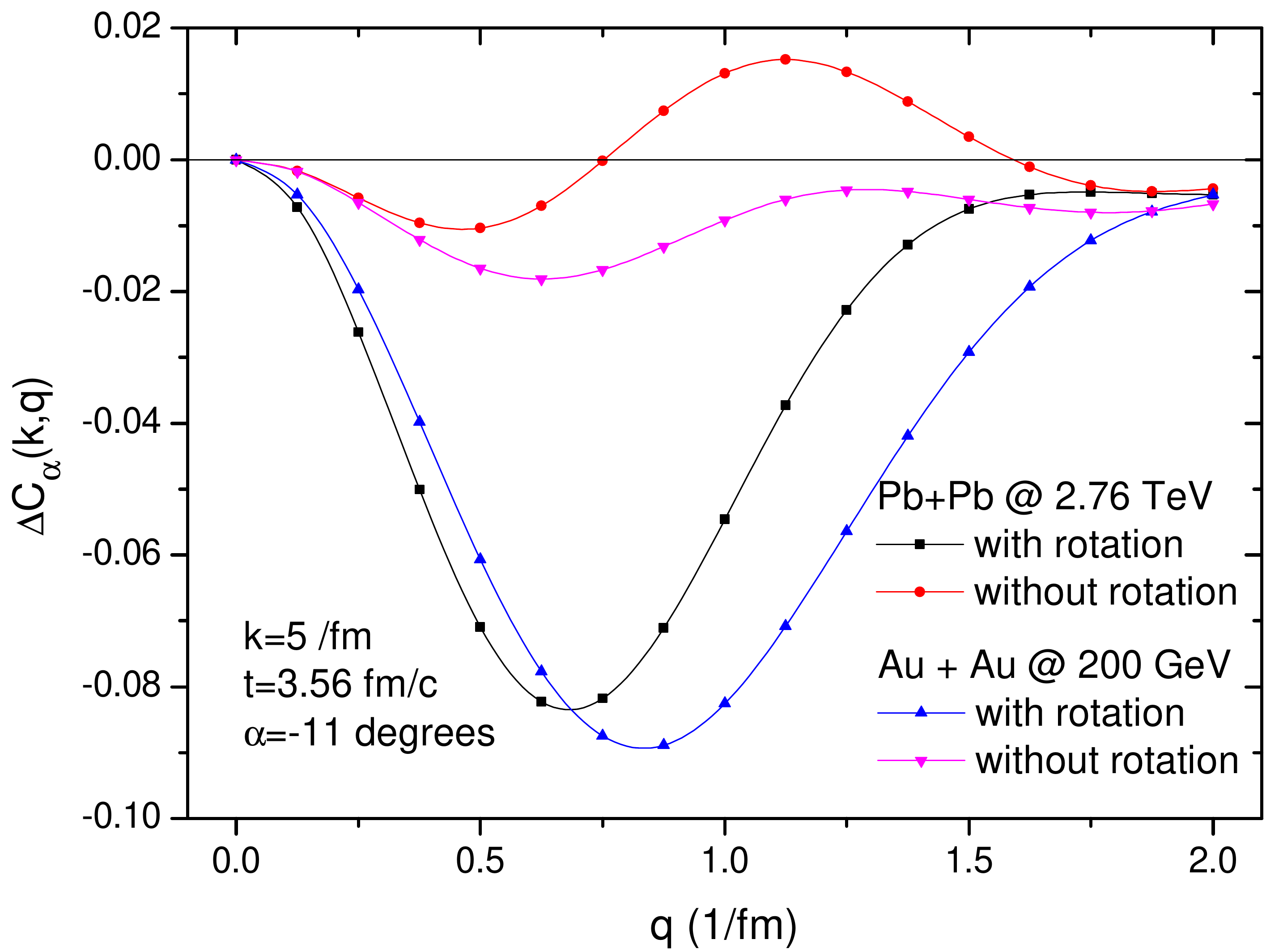}
	\end{center}
	\vskip -0.3cm
	\caption{(color online)
		The differential correlation function (DCF) with and without rotation in the
		reference frames, deflected by the angle $\alpha$, where the rotation-less
		DCF is vanishing or minimal. In this frame the DCF of the original, rotating
		configuration indicates the effect of the rotation only. 
		The amplitude of the DCF 
		of the original rotating configuration doubles for the higher energy (higher
		angular momentum) collision. From ref. \cite{CVW2014}.
	}
	\label{Fig-4}
\end{figure}

Finally we separated the effect of the rotation by finding the
symmetry angle where the rotation-less configuration yields vanishing 
or minimal DCF for a given transverse momentum $k$. 
This could be done in the theoretical model.
We did this for two different energies, Pb+Pb / Au+Au
at $\sqrt{s_{NN}}= 2.36 / 0.2$ TeV respectively, 
while all other parameters of the collision were the same. 
The deflection angle of the symmetry axis was $\alpha =-11/-8$ 
degrees\footnote{The negative angles are arising from the fact that
our model calculations predict rotation, with a peak rotated forward
\cite{hydro1}.}
respectively. In these deflected frames we evaluated the DCF for the
original, rotating configurations, which are shown in  Fig.\,\ref{Fig-4}.
This provides an excellent measure of the rotation.

\section{Summary}


We show that two particle correlation measurements can be 
sensitive to the rotation of the emitting system.
The analysed model calculations show that the Differential HBT analysis 
can give a good quantitative measure of the
rotation in the reaction plane of a heavy ion collision.



\end{document}